\documentstyle[]{article}

\newcommand\mathC{{\mkern1mu\raise2.2pt\hbox{$\scriptscriptstyle|$}
        {\mkern-7mu\rm C}}}
\newcommand{\mathR}{{\rm I\! R}}       

\renewcommand\mathR{{\rm I\! R}}

\newcommand\unit{{\rm 1\kern-3.2pt I}}

\renewcommand\a{\alpha}                         

\newcommand\e{\epsilon}

\begin{document}
\title{ The Barbero connection and its relation to the
histories connection formalism without gauge fixing}
\author{ Ntina Savvidou \footnote{ntina@imperial.ac.uk} \\
 {\small  Theoretical Physics Group, Imperial College, SW7 2BZ,
London, UK}}

\maketitle

\begin{abstract}

We present a histories version of the connection formalism of
general relativity. Such an approach introduces a spacetime
description---a characteristic feature of the histories
approach---and we discuss the extent to which the usual loop
variables are compatible with a spacetime description. In
particular, we discuss the definability of the Barbero connection
without any gauge fixing. Although it is not the pullback of a
spacetime connection onto the three-surface and it does not have a
natural spacetime interpretation, this does not mean that the
Barbero connection is not suitable variable for quantisation; it
appears naturally in the formalism even in absence of gauge fixing.
It may be employed therefore, to define loop variables similar to
those employed in loop quantum gravity. However, the loop algebra
would have to be augmented by the introduction of additional
variables.

\end{abstract}

\renewcommand {\thesection}{\arabic{section}}
 \renewcommand {\theequation}{\thesection.\arabic{equation}}
\let \ssection = \section
\renewcommand{\section}{\setcounter{equation}{0} \ssection}

\section{Introduction}
In this paper we study the connection formalism for general
relativity within the context of the histories framework. This is a
continuation of previous work on the histories description of
general relativity, in terms of its geometrodynamic variables
\cite{Sav03a, Sav03b}. The main motivation  comes from the fact that
the histories description provides a natural spacetime description
of canonical general relativity and fully incorporates the
fundamental principle of general covariance. The connection
formalism provides the background for the quantisation in terms of
the loop variables: it is thus  the first step in an attempt to
provide a histories quantisation scheme for gravity that follows the
main idea of the loop quantum gravity programme, namely that the
basic kinematical variables of the quantum theory have support on
one-dimensional objects--loops or graphs. This scheme would combine
the spacetime perspective and emphasis on general covariance of the
histories framework with the technical facilities in the quantum
treatment of the constraints, provided by the loop quantum gravity
programme.

The basic object in the histories description is the notion of a
history. This corresponds to the specification of information about
the state of a physical system at different moments of time. It
arises from the consistent histories approach to quantum theory
developed by Griffiths \cite{Gri84}, Omn\'es \cite{Omn8894},
Gell-Mann and Hartle \cite{GeHa9093, Har93a}. The specific version
of the histories approach we describe here is known as the Histories
Projection Operator (HPO) approach, which possesses two distinctive
features. First, a history is a {\em temporally extended} object
that it is represented quantum mechanically by a {\em single}
projection operator on a suitably constructed Hilbert space
\cite{I94, IL94}. Second, the theory possesses a novel temporal
structure, since time is implemented by {\em two distinct
parameters}, one of which refers to the kinematical set-up of the
theory, while the other refers solely to its dynamical behavior
\cite{Sav99a}. At the classical level the two parameters coincide
for all histories that correspond to the solutions of the equations
of motion.

The features above imply that the HPO theory is endowed with a rich
kinematical structure. In the case of general relativity this
results to the fact that the different `canonical' descriptions of
the theory, that correspond to different choices of spacelike
foliation coexist in the space of histories and may be related by a
properly defined transformation \cite{Sav03b}. This allows the
preservation of the spacetime description of the theory, even if one
chooses to work with canonical variables.

The histories theory provides a formalism that allows the
incorporation of other theories, enriching them with a purely
spacetime kinematical description, while preserving the main
features of their dynamical behavior. Moreover, the histories theory
may be developed in a quantisation scheme on its own right, which is
expected to be characterised by two important features: first the
preservation of the full spacetime description even at the quantum
level and second, the quantum mechanical treatment of the full
Lorentzian metric--and not only of its spatial components or its
perturbations around a fixed background.

The technical problem in a histories-based quantisation of general
relativity  is the rigorous implementation of the dynamics by a
history analogue of the Hamiltonian constraint operator: as in the
standard canonical theory, the classical expression is
non-quadratic---indeed non-polynomial---in the field variables, and
so the construction of an operator for the Hamiltonian constraint
seems a hopeless task using conventional methods. For this reason,
the most promising strategy would be to exploit the basic ideas of
loop quantum gravity, which has made the greatest progress in the
construction of such an operator \cite{loopreview}.

The mainline approach towards loop quantum gravity has as a starting
point the formulation of general relativity in terms of the Barbero
connection \cite{Barb}: this connection defines the holonomy
algebra, the representation theory of which provides the
construction of the kinematical Hilbert space in loop quantum
gravity.

The Barbero connection can either be obtained from a canonical
transformation in a space extending the phase space of general
relativity, or from a covariant Lagrangian, written by Holst
\cite{Holst}, through a Legendre transform. The latter procedure,
however, involves a choice of the temporal gauge, and for this
reason the very definition of the Barbero connection is
gauge-dependent. A consequence of gauge-dependence is the loss of
the spacetime background independence of the theory.

The most important benefit provided by the introduction of the
Barbero connection is that it is a connection with respect to the
SU(2) group. This fact makes much easier the study of the
representation of the loop algebra, because SU(2) is a compact group
that possesses a unique normalised Haar measure.

It would clearly be a benefit to the histories formalism if an
object with the properties of the Barbero connection could be
identified without compromising the gauge invariance---and hence the
spacetime covariance---of the theory. We demonstrate that this is
indeed the case in section 4.

The structure of this paper is the following. In Section 2 we
provide a brief summary of the histories theory applied to general
relativity---for details see \cite{Sav03a, Sav03b}.

In Section 3 we develop the histories description for general
relativity expressed in terms of the connection and tetrad
variables, using the Holst Lagrangian.

In Section 4 we examine the definability of the Barbero connection
in absence of gauge fixing and comment on possible approaches to
quantisation. In the last section we summarise and discuss our
results.

\section{Background: A histories version of general relativity}

\paragraph{The basic structure of the HPO histories theory.}
In the consistent-histories theory a history is defined as a
sequence of time-ordered propositions about properties of a physical
system, each one represented by a projection operator. When a
certain `decoherence condition' is satisfied by a set of histories,
the elements of this set can be given probabilities. The probability
information of the theory is encoded in the decoherence functional,
a complex function of pairs of histories.

In the HPO approach of the histories theory, propositions about the
histories of a system are represented by projection operators on a
new, `history' Hilbert space. One may understand the history Hilbert
space ${\mathcal V}$ in terms of the representations of the `history
group' \cite{IL95,ILSS98}---in elementary systems this is the
history analogue of the canonical group. For example, for the simple
case of a point particle moving on a line, the history group for a
continuous time parameter $t$ is described by the history
commutation relations
 \vspace*{-0.3cm}
 \begin{eqnarray}
 {[}x_t,x_{t'}]&=&0 = {[}p_t,p_{t'}]  \label{HA1} \\
 {[}x_t,p_{t'} ] &=& i \hbar \delta(t-t') , \label{HA3}
 \end{eqnarray}
where the spectral projectors of the (Schr\"odinger picture)
operators $x_t$ and $p_t$ represent the values of position and
momentum respectively at time $t$. This particular history algebra
is equivalent to the algebra of a $1+1$-dimensional quantum field
theory, and hence techniques from quantum field theory can be used
in the study of the history Hilbert space. This was done
successfully in \cite{ILSS98}, where we showed that the physically
appropriate representation can be uniquely constructed by demanding
the existence of a time-averaged Hamiltonian operator $H:=\int
dt\,H_t$.

The study of continuous-time transformations led a very important
result for the temporal structure of the theory: there exist {\em
two\/} distinct types of time transformation. One refers to time as
it appears in temporal logic---the $t$-label in Eqs.\
 (\ref{HA1}--\ref{HA3}). The other refers to time as it appears in
 the implementation of dynamical laws---the label $s$ in the
 `history Heisenberg picture' operator $ x_t(s):=e^{isH}x_t
 e^{-isH}$. The definition of these two distinct operators,
 implementing time transformations, signified the {\em distinction
 between the kinematics and the dynamics of the theory\/}. More
 important, for any specific physical system these two
 transformations are intertwined by the definition of the {\em action
 operator}---a quantum analogue of the classical action functional.

In the classical histories theory, the basic mathematical entity is
the space of differentiable paths $\Pi  = \{ \gamma \mid\gamma  :
\mathR \rightarrow \Gamma \}$, taking their value in the space of
single-time classical phase space $\Gamma$. The key idea in this new
approach to classical histories is contained in the symplectic
structure on this space of temporal paths $t\rightarrow (x_t , p_t
)$ corresponding to the following Poisson brackets
\begin{eqnarray*}
  \{x_t , x_{t^{\prime}} \} &=& 0 = \{p_t , p_{t^{\prime}} \}\\
  \{x_t , p_{t^{\prime}} \} &=& \delta (t , t^{\prime}),
\end{eqnarray*}
 where
\begin{eqnarray*}
 x_t : \Pi & \rightarrow & \mathR  \\
 \gamma & \mapsto & x_t (\gamma) := x( \gamma (t)).
\end{eqnarray*}
Analogous to the quantum case, there are generators for two types of
time transformation: one associated with classical temporal logic,
and one with classical dynamics.

The classical Hamilton equations may be written in terms of the
Liouville function $V$ and the smeared Hamiltonian function $H$,
which are the classical analogues of the corresponding operators we
defined for the quantum theory
\begin{eqnarray*}
 \{ F_t , V \}_{\Pi} (\gamma_{cl}) = \{ F_t , H \}_{\Pi}
 (\gamma_{cl}),
\end{eqnarray*}
 where $V(\gamma) := \int dt p_t \dot{x}_t$, and $\{F_t,
 V\}=\dot{F}_t$.
One significant feature is that the paths corresponding to solutions
of the classical equations of motion are determined by the
requirement that they remain invariant under the symplectic
transformations generated by the action $S$, for all functions $F_t$
\begin{equation}
  \{\, F \,, S\, \}_{\Pi} ({\gamma}_{cl}) = 0, \label{lap}
\end{equation}
where $ S = V-H $. The Eq.\ (\ref{lap}) is essentially the histories
analogue of the least action principle.

\paragraph{Spacetime description of histories general relativity
theory.} A significant result emerged from the histories general
relativity theory: it is possible to develop a histories theory for
quantising the full spacetime metric of classical
gravity\cite{Sav03b}.

In this context, a `covariant' description of the histories gravity
theory has been developed \cite{Sav03a,Sav03b}, in terms of a
Lorentzian metric $g$, and its `conjugate momentum' tensor $\pi$, on
a spacetime manifold $M$, with topology $\Sigma \times \mathR$. The
history space has as elements the pairs $( g_{\mu\nu} ,
\pi^{\kappa\lambda})$, and it is equipped with the covariant
symplectic form
\begin{eqnarray}
\Omega_{cov} = \int d^4 X \; \delta \pi^{\mu \nu}(X) \wedge \delta
g_{\mu \nu} (X).
\end{eqnarray}

\paragraph{The spacetime diffeomorphisms group $\rm Diff(M)$.}
The relation between the spacetime diffeomorphisms algebra, and the
Dirac constraint algebra has long been an important matter for
discussion in quantum gravity. In this new construction, the two
algebras appear together for the first time: the history theory
contains a representation of both the spacetime diffeomorphism group
{\em and\/} the Dirac algebra of constraints of the canonical
theory. Indeed, for each vector field $W$ on the spacetime manifold
$M$, the `Liouville' function is defined as
\begin{eqnarray} V_W:=\int \!d^4X
\,\pi^{\mu\nu}(X)\,{\cal L}_W g_{\mu\nu}(X), \label{V_W}
\end{eqnarray}
where ${\cal L}_W$ denotes the Lie derivative with respect to $W$.
The functions $V_W$, satisfy the Lie algebra of the spacetime
diffeomorphisms group $\;[ V_{W_1}\,, V_{W_2}] = V_{ [ W_1 , W_2 ]}
\;$, for all spacetime vector fields $W_1 , W_2$.

\paragraph{Spacetime and canonical descriptions.}
In the standard canonical formalism we introduce a spacelike
foliation ${\cal E}:\mathR\times\Sigma\rightarrow M$ on $M$, with
respect to a fixed Lorentzian four-metric $g$. However a foliation
cannot be spacelike with respect to \textsl{all} metrics $g$ and in
general, for an arbitrary metric $g$ the pullback of a metric
${\cal{E}}^* g$ \textsl{is not} a Riemannian metric on $\Sigma$. The
notion of `spacelike' has \textsl{no a priori} meaning in a theory
of quantum gravity, in which the metric is a non-deterministic
dynamical variable. In absence of deterministic dynamics, the
relation between canonical and covariant variables appears rather
puzzling.

In histories theory this problem is addressed by introducing the
notion of a \textsl{metric dependent} foliation ${\cal{E}}[g]$. This
is defined as a map ${\cal{E}}: {\textsl{LRiem}(M)}\mapsto
{\textsl{Fol}{M}} $, that assigns to each Lorentzian metric $g$ a
foliation that is \textsl{always} spacelike with respect to that
metric; ${\textsl{Fol}{M}} $ is the space of foliations on $M$.

We then use the metric dependent foliation ${\cal{E}}[g]$ to define
the canonical decomposition of the metric $g$ with respect to the
canonical three-metric $h_{ij}$, the lapse function $N$ and the
shift vector $N^i$. For example, the three-metric $h_{ij}$ reads
\begin{eqnarray}
 h_{ij}(t,\underline{x})\!\!\! &:=&\!\!\!
 {\cal{E}}^{\mu}_{,i}(t,\underline{x};g]\,
 {\cal{E}}^{\nu}_{,j}(t,\underline{x};g]\,
 g_{\mu\nu}({\cal{E}}(t,\underline{x};g]).
\end{eqnarray}
Defined in this way $h_{ij}$ is \emph{always} a Riemannian metric,
with the correct signature. Hence, the $3+1$ decomposition preserves
the spacetime character of the canonical variables. In the histories
theory therefore, \emph{the $3+1$ decomposition preserves the
spacetime character of the canonical variables\/}, a feature that we
may expect to hold in a theory of quantum gravity.

One may therefore employ histories of canonical variables as
coordinates on the space  $\Pi^{cov}= T^{*}{\rm LRiem}(M)$. We thus
obtain the history version of the canonical Poisson brackets from
the covariant ones, and we can write the history analogue of the
canonical constraints. The canonical description leads naturally to
a one-parameter family of super-hamiltonians
$t\mapsto{\cal{H}}_{\bot} ( t, \underline{x})$ and super-momenta
$t\mapsto {\cal{H}}_i ( t, \underline{x} )$\cite{Sav03a,Sav03b}.

\paragraph{Invariance transformations.}

The introduction of the \textsl{equivariance condition} leads to an
explicit mathematical relation between the $\rm Diff(M)$ group and
the canonical constraints. This condition follows from the
requirement of general covariance, namely that the description of
the theory ought to be invariant under changes of coordinate
systems, implemented by spacetime diffeomorphisms. Loosely speaking,
the equivariance condition makes it possible that the foliation
functional `looks the same' in all coordinate systems.

A metric-dependent foliation functional ${\cal E}:{\rm
LRiem}(M)\rightarrow {\rm Fol}(M)$ is defined as an
\textsl{equivariant foliation} if it satisfies the mathematical
condition
\begin{eqnarray}
{\cal E}[f^*g]=f^{-1}\circ {\cal E}[g],
\label{Def:EquivariantFoliation2} \nonumber
\end{eqnarray}
for all Lorentzian metrics $g$ and $ f\in{\rm Diff}(M)$.

Hence, if we perform a change of the coordinate system of the theory
under a spacetime diffeomorphism, then the expressions of the
objects defined in it will change, and so the foliation functional
${\cal E}[g]$ and the four-metric $g$ will also change. Then, the
change of the foliation due to the change of the coordinate system
must be compensated by the change due to its functional dependence
on the metric $g$. This is essentially the \emph{passive
interpretation} of the spacetime diffeomorphisms.

The equivariance condition manifests a striking result: the action
of the spacetime diffeomorphisms group $\rm Diff(M)$
\textsl{preserves} the set of the constraints, in the sense that it
transforms a constraint into another of the same type but of
different argument. Hence, the choice of an equivariance foliation
implements that histories canonical field variables related by
diffeomorphisms are physically equivalent.

Furthermore, this result means also that, the group $\rm Diff(M)$ is
represented in the space of the true degrees of freedom, the reduced
phase space. Hence, in the histories theory the requirement of the
physical equivalence of different choices of time direction is
satisfied by means of the equivariance condition.

\paragraph{Reduced state space.}
We define the history constraint surface $C_h = \{ t \mapsto C, t\in
\mathR\}$, as the space of maps from the real line to the
single-time constraint surface $C$ of canonical general relativity.

The history reduced state space is obtained as the quotient of the
history constraint surface $C_h$, with respect to the action of the
constraints, i.e., it is the space of orbits on $C_h$ arising from
the action of the constraints.

The histories version of the Hamiltonian constraint is defined as
$H_{\kappa} = \int \! dt\, \kappa (t) h_t$, where $h_t$ is
first-class constraint. For all values of the smearing function
$\kappa (t)$, the history Hamiltonian constraint $H_{\kappa}$
generates canonical transformations on the history constraint
surface.

The history reduced state space $\Pi_{red}= \{ t\mapsto
\Gamma_{red}, t\in\mathR\}$, is a symplectic manifold that can be
identified with the space of paths on the canonical reduced state
space $\Gamma_{red}$.

It has been proved \cite{Sav03b} that the histories reduced state
space is \textsl{identical} with the space of paths on the canonical
reduced state space. Hence, the time parameter $t$ also exists on
$\Pi_{red}$, and the notion of \emph{time ordering\/} remains on the
space of the true degrees of freedom $\Pi_{red}$. This last result
is in contrast to the standard canonical theory where there exists
ambiguity with respect to the notion of time after reduction.

The phase space action functional $S$ commutes weakly with the
constraints, so it can be projected on the histories reduced state
space. It then determines the equations of motion, as we have shown
in the theory of classical histories \cite{Sav99a}.

A function on the full state space $ \Pi $ may be considered to be a
physical observable ({\em i.e.,\/} to be projectable into a function
on $ {\Pi}_{red}$), if it commutes with the constraints on the
constraint surface.

The equations of motion are the paths on the phase space that remain
invariant under the symplectic transformations generated by the
projected action
\begin{eqnarray} \{ \tilde S
, F_t \}\,(\gamma_{cl}) = 0\,, \nonumber
\end{eqnarray}
where $F_t$ is constant in $t$ and $\tilde{S}$ is the action
projected on $\Pi_{red}$.

The canonical action functional $S$ is also diffeomorphic-invariant
\begin{equation}
 \{ V_W , S \} = 0.
\end{equation}
Therefore, the dynamics of the histories theory is invariant under
the group of spacetime diffeomorphisms.

We can distinguish the paths corresponding to the classical
equations of motion by the condition
\begin{equation}
 \{ F , \tilde{V}\}_{{\gamma}_{cl}} = 0 , \label{FV}
\end{equation}
where $F$ is a functional of the field variables, and
${{\gamma}_{cl}}$ is a solution to the equations of motion.

In standard canonical theory, the elements of the reduced state
space are all solutions to the classical equations of motion. In
histories canonical theory, however, an element of the reduced state
space is a solution to the classical equations of motion only if it
also satisfies the condition Eq.\ (\ref{FV}). The reason for this is
that the histories reduced state space ${\Pi}_{red}$ contains a much
larger number of paths (essentially all paths on ${\Gamma}_{red}$ ).

\section{The histories theory of the connection formalism
for general relativity}

\subsection{The covariant description}

The building block of the basic variables employed in loop quantum
gravity is a pair of conjugate variables, consisting of a densitised
triad $\tilde{E}_a^i$ and an $SU(2)$ three-connection $A_a^i$.  The
description of general relativity in terms of these variables can be
obtained from two different procedures: one may either enlarge the
geometrodynamical state space and identify the relevant variables
through a suitable canonical transformation; or one may start from a
spacetime action \cite{Holst} by following the usual Dirac-Bergmann
theory of constraints (together with partial gauge fixing).

The histories approach highlights the spacetime aspects of general
relativity, hence it is the latter procedure that we shall adopt. As
starting point we consider the Holst action
\begin{eqnarray}
S[E,\omega] = \int d^4 X (E) E^{\mu}_I E^{\nu}_J (\Omega_{\mu
\nu}^{IJ} - \!\beta\!^{-1} {}^*\!\Omega_{\mu \nu}^{IJ}),
\label{Holst}
\end{eqnarray}
where $\beta$ is known as the Immirzi parameter \cite{Immi97}, while
\begin{eqnarray}
\Omega_{\mu \nu}^{IJ} = \partial_{\mu} \omega+n^{IJ} -
\partial_{\nu} \omega_{\mu}^{IJ} + [\omega_{\mu},
\omega_{\nu}]^{IJ},
\end{eqnarray}
and ${}^*\Omega_{\mu \nu}^{IJ}: = \frac{1}{2} \epsilon^{IJ}{}{}_{KL}
\Omega_{\mu \nu}^{KL}$.

The action (\ref{Holst}) is a function on the space $\Lambda$ of
configurational histories, that consists of tetrads $E_{\mu}^I$, and
Lorentz-connections $\omega_{\mu}^{IJ}$, on a four-manifold $M$.
Note that $\Lambda$ carries no symplectic structure.

The space $\Pi_{cov}:=T^*\Lambda$ of phase space histories is the
cotangent bundle of $\Lambda$. It consists of variables $(E_{\mu}^I,
\omega_{\mu}^{IJ}, {}^{E\!}\pi^{\mu}_{IJ}, {}^{\omega
\!}\pi_{IJ}^{\mu})$ and it is equipped with the covariant histories
symplectic form
\begin{eqnarray}
\Omega = \int d^4 X \left[  \delta {}^{E\!}\pi^{\mu}_{I}\wedge
\delta E_{\mu}^I + \delta {}^{\omega \!}\pi_{IJ}^{\mu}\wedge \delta
\omega_{\mu}^{IJ} \right].
\end{eqnarray}

We define functions $V_W$ on $\Pi_{cov}$,
\begin{eqnarray}
V_W = \int d^4 X \; [{}^{E \!}\pi^{\mu}_{I}{\cal L}_W E_{\mu}^I +
{}^{\omega \!}\pi_{IJ}^{\mu}{\cal L}_W \omega_{\mu}^{IJ}],
\end{eqnarray}
 where $W$ is a vector field on $M$ and ${\cal L}$ denotes the Lie
 derivative. The functions $V_W$ correspond to a symplectic action of
 the group $Diff(M)$ on $\Pi_{cov}$ since they satisfy the Lie
 algebra of the $Diff(M)$ group
 \begin{eqnarray}
\{V_{W_1}, V_{W_2} \} = V_{[W_1,W_2]}.
 \end{eqnarray}

 It is easy to check that $V_W$ generates spacetime diffeomorphisms
 upon the canonical variables, for example
 \begin{eqnarray}
\{V_W, E_{\mu}^I \} = - {\cal L}_W E_{\mu}^I \nonumber \\
\{V_W, \omega_{\mu}^{IJ} \} = - {\cal L}_W \omega_{\mu}^{IJ}.
 \end{eqnarray}

\subsection{The $3+1$-decomposition}

Similarly to the case of the metric variables described in Section
2, the space $\Pi_{cov}$ of tetrads and Lorentz connections is fully
compatible with the 3+1 decomposition of the space $\Lambda$ and
hence it can incorporate the description of histories of canonical
variables.

The first step in a 3+1 decomposition involves the specification of
a spacelike foliation. In order to preserve the spacelike character
of the canonical description it is necessary for this foliation to
be a function of the 4-metric $g$, as explained in Sec. 2--see
\cite{Sav03a} for details. The metric is defined on $\Pi_{cov}$ as a
function of the tetrad variables
\begin{eqnarray}
g_{\mu \nu} = \eta_{IJ} E^I_{\mu} E^J_{\nu}.
\end{eqnarray}
Hence, for each specific Lorentzian metric $g$ we choose a spacelike
foliation, ${\cal E}:\mathR\times\Sigma\rightarrow M$, with an
associated family of spacelike embeddings ${\cal
E}_t:\Sigma\rightarrow M$, $t\in\mathR$. Given a foliation
functional ${\cal E}$, we can define the normal unit timelike vector
to the foliation $n^{\mu}(X;g]$, the vector field $t^{\mu}(X) =
\dot{{\cal E}}^{\mu}({\cal E}^{-1}(X))$, which specifies our
direction of time and the `projector' fields ${\cal E}^{\mu}_{,a} =
\frac{\partial {\cal E}^{\mu}}{\partial x^a}$.

The pull-back of the tetrad and the four-connection on each
spacelike 3-surface $\Sigma_t = {\cal E}_t^{-1}(M)$ are defined as
\begin{eqnarray}
{}^3E_a^I(t,x) := {\cal E}^{\mu}_{,a} E_{\mu}^I ({\cal E}(t,x))\\
{}^3\omega_a^{IJ}(t, x) = {\cal E}^{\mu}_{,a}
\omega_{\mu}^{IJ}({\cal E}(t,x)) .
\end{eqnarray}
We also define the lapse function $N$ and the shift vector $N_a$ as
\begin{eqnarray}
N(t,x) &=& {\dot{\cal{E}}}^{\mu}\, n_{\mu}({\cal{E}}
 (t,x)) \label{lapse} \\
 N_{a}(t,x) &=& {\cal{E}}^{\mu}_{,a}\,
 {\dot{\cal{E}}}^{\nu} \, g_{\mu\nu}({\cal{E}}(t,x)),
 \label{shift}
\end{eqnarray}
and the projection of the tetrad along $n^{\mu}$ as
\begin{eqnarray}
n^I(t,X) = n^{\mu} E_{\mu}^I({\cal E}(t,x)).
\end{eqnarray}
The latter satisfies the conditions ${}^3E_a^I n^J \eta_{IJ} = 0, \;
n^I n_I = 1$. We also consider the projections of the
four-connection along $t^{\mu}$
\begin{eqnarray}
\alpha^{IJ} = {\dot{\cal{E}}}^{\mu}\, \omega_{\mu}^{IJ}.
\end{eqnarray}
The 3+1 decomposition results to the substitution of the original
set of `covariant' coordinates $E_{\mu}^I, \omega_{\mu}^{IJ}$ on
$\Lambda$ with the following ones:  ${}^3E_a^I, N, N_a, n^I,
{}^3\omega_a^{IJ}, \alpha^{IJ}$. The new coordinates are essentially
paths from the axis of time $\mathR$ to the single-time
configuration space  of the theory.

It is very convenient to choose a new set of variables. One of the
aims of this paper is to see how the Barbero connection appears
within a gauge-invariant spacetime description. The Barbero
connection is defined after a specific gauge fixing: choosing the
temporal gauge $n^I = (1,0,0,0)$. For this reason we choose
variables that are convenient for keeping track of this particular
gauge.

Let us denote by $\Lambda^I_J(t,x)$ the $SL(2,{\bf C})$ gauge
transformation that takes the scalar fields $n^I$ to the constant
vector ${}^0 n^I = (1,0,0,0)$. We write,
 \begin{eqnarray}
\Lambda^I_J(t,x) = \delta^I_J + \frac{1}{n\cdot {}^0n - 1} (n^I -
{}^0n^I)(n_J - {}^0n_J). \label{boost}
 \end{eqnarray}

Next, we transform by $\Lambda^{IJ}$ all fields with internal
indices. We define
\begin{eqnarray}
{}^3E_a'^I := \Lambda^I{}_J {}^3E_a^J \\
{}^3\omega'^{IJ}_a := \Lambda^I{}_K {}^3\omega_a^{KL} \Lambda_L{}^J
\\
 \alpha'^{IJ} := \Lambda^I{}_K {}^3\alpha^{KL} \Lambda_L{}^J.
\end{eqnarray}

But ${}^3E^I_a n_I = 0$, hence ${}^{3}E_a'^I$ is of the form $(0,
E_a^1, E_a^2, E_a^3)$. We notice that the variables $E_a^i = (E_a^1,
E_a^2, E_a^3)$ define a field of triads on $\sigma$, since
\begin{eqnarray}
E_a^i E_b^j \delta_{ij} = h_{ab}:= {\cal E}^{\mu}_a {\cal E}^{\nu}_b
g_{\mu \nu}.
\end{eqnarray}
The property above distinguishes the variables we employ here, from
the variables $E_a^i$ of Ref.\cite{BeS01}, which become triads only
after the imposition of the temporal gauge.

The form of the Holst Lagrangian, suggests that the following
combinations of the connection variables are the most convenient

\begin{eqnarray}
\omega_a^i :&=& {}^3\omega_a'^{0i} - \frac{1}{2\beta} \epsilon^{ijk}
{}^3\omega_a'^{jk} \\
\chi_a^i :&=& \frac{1}{2} \epsilon^{ijk}{}^3 \omega_a'^{jk} +
\frac{1}{\beta}  {}^3\omega_a'^{0i} \\
\rho^i :&=& \a'^{0i} - \frac{1}{2\beta} \epsilon^{ijk} \a'^{jk} \\
\sigma^i :&=& \frac{1}{2} \epsilon^{ijk} \a'^{jk} + \frac{1}{\beta}
\a'^{0i}.
\end{eqnarray}

For further convenience in the description of the constraints we
employ the densitised triad $\tilde{E}^a_i = (E) E^a_i$ and the
densitised lapse $\tilde{N} = N/E$, where $E$ is the square root of
the determinant of the three-metric $h_{ab}$. We also write $n^I =
(\sqrt{1+\xi\cdot \xi}, \xi^i)$.

Hence we have chosen a coordinate system on $\Lambda$, based on the
canonical history variables $(\tilde{E}^a_i, \omega_a^i, \chi_a^i,
\rho^i, \sigma^i, \tilde{N}, N_a, \xi^i)(t,\underline{x})$.  The
change of variables on $\Lambda$ is essentially a contact
transformation. It can therefore be lifted to a symplectic
transformation in $\Pi_{cov} = T^*\Lambda$. Hence the symplectic
form on $\Pi_{cov}$ can be written as
\begin{eqnarray}
\Omega = \int dt \int d^3 x  \left( \delta {}^{E\!}p_a^i \wedge
\delta E_a^i + \delta {}^{\omega \!}p^a_i \wedge \delta \omega_a^i +
\delta
{}^{\chi \!} p^a_i \wedge \delta \chi_a^i \right. \nonumber \\
\left. + \delta {}^{\rho \!}p_i \wedge \delta \rho^i + \delta
{}^{\sigma \!}p_i \wedge \delta \sigma^i  + \delta {}^{\tilde{N}
\!}p \wedge \delta \tilde{N} + \delta {}^{N \!} p^a \wedge \delta
N_a + \delta {}^{\xi \!} p_i \wedge \delta \xi^i \right).
\label{sympl3+1}
\end{eqnarray}

The momenta in (\ref{sympl3+1}) can be expressed as functions of the
covariant momenta and configuration variables. The explicit
relations can be obtained  by substituting  $E_{\mu}^I$ and
$\omega_{\mu}^{IJ}$ by the corresponding canonical variables in the
expression for the symplectic potential $\Theta_{cov} = \int d^4X
[{}^E\pi^{\mu}_{I} \delta E_{\mu}^I +{}^{\omega}\pi_{IJ}^{\mu}\wedge
\delta \omega_{\mu}^{IJ}$, by the canonical variables. For this
purpose we use the equations
\begin{eqnarray}
E_{\mu}^I &=& {\cal E}_{\mu}^a \; {}^3E_a^I + n_{\mu} n^I \\
\omega_{\mu}^{IJ} &=& ({\cal E}_{\mu}^a - N^{-1} n_{\mu} N^a) \;
{}^3\omega_a^{IJ} +  N^{-1} n_{\mu} \alpha^{IJ}.
\end{eqnarray}

The expression (\ref{sympl3+1}) demonstrates that the space $\Pi$ is
isomorphic to the space of paths over the phase space of the
corresponding canonical theory.

\subsection{The constraints}
So far the description is purely kinematical; the only input has
been the choice of variables for the covariant description and the
implementation of the 3+1 splitting. In order to introduce dynamics,
we must consider the constraints corresponding to the Holst action
\cite{Holst}.

In the Appendix we provide details about the Legendre transform of
the Holst Lagrangian. The approach followed in this paper is
different from the one usually encountered in the literature,
because it preserves the relation of the canonical variables to the
original covariant ones\footnote{The usual procedure involves
substituting the second-class constraints that appear in the
Legendre transform (\ref{1sc}, \ref{2sc}) by the single one that is
expressed in terms of the conjugate momentum of ${}^3\omega_a^{IJ}$,
namely $\e^{IJKL} {}^{\omega}p_{IJ}^a {}^{\omega}p_{KL}^b = 0$.
While this method is more convenient for computational purposes, it
hides the immediate relation between the components of the tetrad
and the covariant momentum conjugate to the Lorentz connection.}. We
would not be able otherwise to verify the invariance of the
constraints and the equations of motion under the spacetime
diffeomorphisms action on $\Pi_{cov}$. The other difference of the
Legendre transform we develop in the Appendix is that the canonical
variables we employ have a natural geometric interpretation even in
absence of gauge fixing.

The constraints restrict the space of possible histories to a
submanifold of $\Pi_{cov}$. We identify the second-class primary
constraints ($30 \times \infty^4$)
\begin{eqnarray}
{}^{\omega}p^a_i - \tilde{E}^a_i = 0 \label{1sc}\\
{}^{\chi}p^a_i = 0 \label{2sc}\\
{}^Ep_a^i = 0 \label{3sc}\\
{}^{\xi}p^l + \e^{ijk}\tilde{E}^a_i[\chi_a^j v^k_l + \omega_a^j
w^k_l] = 0 \label{4sc},
\end{eqnarray}
the ($6 \times \infty^4$) second class secondary constraints (for
the full expression of the functions $f^{ij}$ see the Appendix)
\begin{eqnarray}
2 \tilde{E}^{a(i} \chi_a^{j)} = f^{ij}(\tilde{E}, \omega, \xi)
 \label{5sc},
\end{eqnarray}
the $(10\times \infty^4$) primary first-class constraints
\begin{eqnarray}
{}^{\tilde{N}}p = 0 \label{pf1}\\
{}^Np^a = 0 \label{pf2} \\
{}^{\rho}p_i = 0 \label{pf3}\\
{}^{\sigma}p_i = 0 \label{pf4},
\end{eqnarray}
and the $(10\times \infty^4$) secondary first-class constraints
\begin{eqnarray}
{\cal F}^k &=& - \tilde{E}^a_i
\e^{ijk}[(1+\beta^{-2})^{-1}(\omega_a^j + \beta^{-1} \chi_a^j)+
v^j_l
\partial_a \xi^l) = 0 \label{fsc1}
\\
 {\cal G}^k &=&
\partial_a \tilde{E}^a_k  - \e^{ijk} \tilde{E}^a_i [(1+\beta^{-2})^{-1}
(\chi_a^j- \beta^{-1} \omega_a^j) +
 w^j_l \partial_a \xi^l] = 0 \label{fsc2} \\
 {\cal H}_a &=& -\tilde{E}^b_i \left[2 \partial_{[a}\omega_{b]}^i -
 \e^{ijk}\omega_b^j v^k_l \partial_a  \xi^l + 2\e^{ijk}\omega_{[a}^j
 \partial_{b]} \xi^l w^k_l \right] \nonumber \\
 &-& (-2 \omega_a^j L^j - 2 (1 + \beta^{-2}) \partial_a \xi^l v^k_l
 L^k) - \chi_a^j {\cal G}_{\nu}^i  \label{fsc3} \\
 {\cal H} &=& \tilde{E}^a_i \tilde{E}^b_j \e^{ijk} [
 (2\partial_{[a}\chi^k_{b]}- (1+\beta^{-2})^{-1} \e^{kmn}
 (\omega_a^m \omega_b^n -\chi_a^m\chi_b^n
\nonumber \\
&+&\beta^{-1}\chi_a^m \omega_b^n + \beta^{-1}\omega_a^m \chi_b^n))
 + 2\e^{kmn}(\omega_{[b}^m v^n_l - \chi_b^m w^n_l)
\partial_{a]} \xi^l ], \label{fsc4}
\end{eqnarray}
where $w^m_n, v^m_n$ are functions of $\xi$ defined in
Eq.(\ref{vw2}) of the Appendix, and $L^k = \frac{1}{2} \e^{ijk}
\tilde{E}^{ai}\chi_a^j$.

The constraints ${\cal H}$, ${\cal H}_a$ are obtained by the
variation of the lapse $N$ and the shift $N_a$ respectively, hence
they correspond to the Hamiltonian and momentum constraint, while
the constraints ${\cal G}^i, {\cal F}^i$ are obtained by the
variation of $\rho_i$ and $\sigma_i$ respectively. The latter
constraints correspond to the $SL(2,{\bf C})$ gauge constraints of
the usual treatment. At this level the algebra of the secondary
first-class constraints is trivial, because the variables
$\tilde{E}_i^a, \omega_a^i$ commute prior to the imposition of the
second-class constraints (\ref{1sc}).

\subsection{The spacetime diffeomorphism invariance}

In order to demonstrate the diffeomorphism invariance of the reduced
state space, we need to restrict our considerations to equivariant
foliation functionals--see Sec. 2. We denote by $A(\cdot , g]$ any
tensor field associated to the foliation functional, that carries a
dependence on the metric $g$. The physical requirement is that the
change of the tensor field $A$ under a diffeomorphism transformation
is compensated by the change due to its functional dependence on
$g$.

For an infinitesimal spacetime diffeomorphism transformation the
equivariance condition (\ref{Def:EquivariantFoliation2}) can be
expressed as
\begin{equation}
 {\cal L}_W A(X ; g] = \int\!\!d^4X^{\prime}\; \frac{\delta A(X;g]}
 {\delta g_{\mu\nu}(X^{\prime})} {\cal L}_W
 g_{\mu\nu}(X^{\prime}) \,.
\end{equation}

The analysis of the diffeomorphism invariance of the constraints
proceeds as in the case of \cite{Sav03b}. The result is identical,
namely that the constraints are preserved by the action of the
diffeomorphisms generated by the $V_W$. It can be verified by
explicit calculation, however the general argument is the following.

The functions of the constraints may be smeared with suitable fields
on $\Sigma \!\times \!\mathR$. For example, the momentum constraint
may be expressed in smeared form as ${\cal H}(M) = \int dt d^3 x
{\cal H}_a(t,x) M^a(t,x)$. In general, the smeared form of a
constraint $F^r$, where $r$ refers to any indices, may be denoted by
the expression $F(M) = \int dt d^3x F^r(t,x) M_r(t,x)$. We
substitute the canonical variables employed in the definition of the
constraints, by their expressions in terms of the covariant objects
$E_{\mu}^I, \omega_{\mu}^{IJ}, {}^E\pi^{\mu}_I,
{}^\omega\pi^{\mu}_{IJ}$. For simplicity, we refer to all the
covariant fields variables as $\phi_A$ (we  only need to distinguish
the tetrad later on).

We write the constraint expressed in terms of the covariant
variables as $\int d^4X \, F^r(X,\phi_A, {\cal E}] \, M_r(X;g]$,
where ${\cal E}$ are any fields defined with reference to the
spacelike foliation (namely ${\cal E}^{\mu}_{,a}$ and $n^{\mu}$. The
smearing fields $M^r$ now depend on $g$, because they are
pull-backed to $M$ using the foliation functional. We then compute
\begin{eqnarray}
\{V_W, F(M) \} = \int d^4 X d^4 X' \left[ \frac{\delta F^r(X, \phi,
{\cal E}) }{\delta \phi_A(X')} {\cal L}_W \phi_A(X')
M^r(X,g] + \right. \nonumber \\
\left. \left( \int d^4X''  \frac{\delta F^r(X, \phi, {\cal E})
}{\delta {\cal E}(X'', g ]} \frac{\delta {\cal E}(X'',g]}{\delta
g_{\mu \nu}(X')} M^r(X, g] \right. \right. \nonumber \\
\left. \left. + F^r(X) \frac{\delta M^r(X,g]}{\delta g_{\mu
\nu}(X')} \right) 2 E_{\mu}^I(X') E_{\nu I}(X')\right].
\end{eqnarray}
Using the equivariance condition we obtain
\begin{eqnarray}
\{V_W, F(M) \} = \int d^4 X {\cal L}_W (F^r(X) M_r(X,g]) \nonumber
\\
+ \int d^4 X F^r(X)\left( -{\cal L}_W M^r +  \int d^4 X'
\frac{\delta M^r(X,g]}{\delta g_{\mu \nu}(X')} 2 E_{\mu}^I(X')
E_{\nu I}(X')
\right) \nonumber \\
:= -F(\delta_W M),
\end{eqnarray}
where $\delta_W M = {\cal L}_W M^r - \int d^4 X' \frac{\delta
M^r(X,g]}{\delta g_{\mu \nu}(X')}  2 E_{\mu}^I(X') E_{\nu I}(X')$,
is an infinitesimal change to $M^r$ due to the total action of the
diffeomorphisms.

Hence,  the action of $V_W$ upon the constraints leaves the set of
constraints invariant. It follows that $V_W$ can be projected to the
histories reduced state space.

The same argument holds for the equations of motion. One may define
the Liouville function on $\Pi_{cov}$ as
 \begin{eqnarray}
V = \int dt \int d^3 x  \left( {}^Ep_a^i  \dot{E}_a^i +
{}^{\omega}p^a_i  \dot{\omega}_a^i +  {}^\chi p^a_i \dot{\chi}_a^i
+ {}^{\rho}p_i \dot{ \rho \!}^i + {}^{\sigma}p_i \dot{ \sigma \!}^i
\right. \nonumber \\
\left. +  {}^{\tilde{N}}p  \dot{\tilde{N}} + {}^N p^a  \dot{N}_a +
 {}^{\xi} p_i \dot{ \xi}^i \right).
 \end{eqnarray}

Since the constraint functions are all ultra-local in $t$, i.e. they
do not involve derivatives with respect to $t$, we obtain
\begin{eqnarray}
\{V, F(M) \} = F(\dot{M}).
\end{eqnarray}
Hence, $V$ commutes weakly with the constraints and thus it can be
projected to the reduced state space. In the reduced state space the
Hamiltonian is zero, and for this reason the projection of $V$ there
coincides with the projection of the action functional. It therefore
generates the solution to the equations of motion: if $\tilde{V}$ is
the projection of $V$ on the reduced history space then the equation
\begin{eqnarray}
\{\tilde{V}, F \}(\gamma_{cl}) = 0
\end{eqnarray}
identifies the classical equations of motion, for any function $F$
on $\Pi_{red}$.

Finally, using arguments similar to the above, it is easy to
demonstrate that $V$ remains invariant under the action of the group
$V_W$ of spacetime diffeomorphisms, provided that the foliation is
equivariant. The derivation is formally identical to that of Ref.
\cite{Sav03b}, where the reader is referred to for details.

\section{The Barbero connection}
The loop variables employed in quantisation are defined in terms of
the Barbero connection, which forms a canonical pair with the triad
on a state space. The latter is constructed after the imposition of
some of the constraints \cite{Holst,BeS01}. To be precise, the
constraints that are imposed on this construction are the
second-class ones and the primary first-class ones, while `half' of
the gauge constraints are removed through the selection of the
temporal gauge.

In this section, we shall consider the issue of the definability of
the Barbero connection without any gauge fixing. This will allow us
to discuss the extent to which the usual loop variables are
compatible with a spacetime description. Note that while we work
within the history formalism for definiteness, the arguments we
present can be immediately translated to the canonical (single-time)
context.

The history space is a symplectic manifold, and for this reason one
may follow the same procedure for implementing the constraints as in
the standard canonical case.

The first step is to impose the second-class constraints. We denote
the resulting manifold as $\Pi_{s.c.}$. By definition (of the
second-class constraints concept) the symplectic form on
$\Pi_{s.c.}$ is non-degenerate. The constraints
(\ref{1sc}--\ref{3sc}) imply that the triads and the
$\omega_a^i$-part of the four-connection $\omega_{\mu}^{IJ}$ form a
pair of conjugate variables. The symplectic form on $\Pi_{s.c.}$ is

\begin{eqnarray}
\Omega_{s.c.} = \int dt \int d^3 x [\delta \tilde{E}^a_i \wedge
\delta \omega_a^i + \delta {}^\xi p_i \wedge \delta \xi^i + \delta
{}^{\rho}p_i \wedge
\delta \rho^i + \delta {}^{\sigma}p_i \wedge \delta \sigma^i
\nonumber \\
 + \delta {}^{\tilde{N}}p \wedge \delta \tilde{N} + \delta {}^N
p^a \wedge \delta N_a].
\end{eqnarray}

The constraints (\ref{4sc}, \ref{5sc}) determine the components
$\chi_a^i$ in terms of the other variables. The primary first-class
constraints commute with the secondary first-class ones, hence they
can be imposed separately. By imposing the former and then excising
the degenerate directions, we construct the space $\Pi_1$. The
latter is spanned by the variables $\tilde{E}, \omega, \xi,
{}^{\xi}p$ and it carries the symplectic form
\begin{eqnarray}
\Omega_1 = \int dt \int d^3 x [\delta \tilde{E}^a_i \wedge \delta
\omega_a^i + \delta {}^\xi p_i \wedge \delta \xi^i].
\end{eqnarray}
First, let us recall that the Barbero connection is defined on a
phase space, which is obtained by the choice of the temporal gauge
$\xi^i =0$ \cite{Holst}. Gauge fixing allows the solution of one
half of the gauge constraints. This results to the determination of
${}^{\xi \!}p$ in terms of the remaining canonical variables. The
object $\omega_a^i$ transforms then as an SU(2) connection under the
remaining gauge constraints: this is the Barbero connection.

However, one may follow a different direction that allows the
identification of the Barbero connection on the space $\Pi_1$,
without any gauge fixing. For this purpose, we find a suitable
variable that transforms as an $SU(2)$ connection under a
combination of the gauge constraints. We notice that the combination
$\beta {\cal F}- {\cal G}$  leads to the Gauss-law constraint
\begin{eqnarray}
{\cal G}^k_{Gauss} = \partial_a \tilde{E}^a_k + \beta \e^{ijk}
\tilde{E}^a_i (\omega_a^j - \beta^{-1} w^j_l \partial_a \xi^l +
v^j_l \partial_a \xi^l) = 0   \label{gaugerot},
\end{eqnarray}
where $v^j_l, w^j_l$ are given by Eqs. (\ref{vw2}) in the Appendix.

From  Eq. (\ref{gaugerot}) we notice that the object
\begin{eqnarray}
A_a^i  = \omega_a^i - \beta^{-1} w^i_l \partial_a \xi^l + v^i_l
\partial_a \xi^l
\end{eqnarray}
transforms as an $SU(2)$ connection\footnote{We should recall that
at this level (in absence of spinors and prior to quantisation) the
gauge group is $SO(3)$ rather than $SU(2)$, but we ignore this
distinction throughout this paper.} under ${\cal G}^i_{Gauss}$, it
is conjugate to the triad and coincides with the Barbero connection
on the gauge fixing surface $\xi^i = 0$. Hence, it is the proper
pull-back of the Barbero connection on $\Pi_1$.

Using Eq. (\ref{4sc}) we bring  the  constraints ${\cal F}^k = 0$ on
$\Pi_1$ into the form ${\cal F'}^k = 0$, where
\begin{eqnarray}
{\cal F'}^k = {}^{\xi}p^k + \tilde{E}^a_i \e^{ijk}[(w^k_l - \beta
v^k_l) \omega_a^j - \beta (1+ \beta^{-2}) v^j_m v^k_l \partial_a
\xi^m]. \label{boostgauge}
\end{eqnarray}

The implementation of the constraints (\ref{boostgauge}) should
allow us to get rid of the $\xi$ and  ${}^{\xi}p$ variables and
descend to the Barbero phase space. This, however, cannot be done in
a gauge-invariant way. The reason is that the submanifold ${\cal
F'}^i =0$ is not preserved by the action of the remaining
constraints---in particular the ${\cal G}^k_{Gauss}$ constraint.
Moreover, the constraint ${\cal F'}$ leaves neither the triad, nor
the Barbero connection invariant. A {\em partial} implementation of
the constraints--while algebraically possible through gauge
fixing--is geometrically inadmissible at this level.

In conclusion, the Barbero connection is well-defined on the space
$\Pi_1$, which is obtained from the imposition of the second-class
and the primary first-class constraints. The Barbero connection
corresponds to a specific combination $A_a^i$ of the system's
variables. This combination transforms as an SU(2) connection under
a specific combination of the $SL(2,{\bf C})$ gauge constraints. It
is however not possible to reduce the system further and go to a
phase space that only contains the connections and the triads,
unless one employs gauge-fixing. This implies that the basic
variables employed in the loop quantum gravity--based as they are on
the Barbero connection--are gauge dependent. The same holds for the
remaining constraints on the Barbero phase space; their specific
form is gauge dependent and for this reason they cannot  be  related
to spacetime objects. Indeed, in the history context the action of
the diffeomorphism group descends on $\Pi_1$, but it cannot descend
to the Barbero phase space, because the introduction of a
gauge-fixing is an additional 'external' structure that violates the
background independence of the theory. The starting point of the
loop quantisation is therefore gauge-dependent and does not have a
straightforward spacetime description.

However, the loop variables may be obtained from a purely canonical
procedure, in which the issue of gauge-dependence may not arise. One
starts from the geometrodynamical phase space (usually after the
implementation of the primary constraints), extends it to include an
$SU(2)$ connection and performs a canonical transformation on the
extended space. This leads essentially to the Barbero variables.
Still, even if the problem of gauge dependence can be avoided, the
problem of spacetime covariance remains:  the resulting canonical
theory (and hence the quantisation scheme) is not spacetime
covariant, because the extension cannot be brought into a
correspondence with a Lagrangian action. One may argue that
spacetime covariance needs not be a fundamental symmetry of a
quantum gravity theory and that it only arises  at the classical
limit.

However, in this paper we contend that a quantisation along the loop
quantum gravity lines is possible, if one starts from either the
full covariant phase space $\Pi_{cov}$, or from the intermediate
space $\Pi_1$ (or its analogues in the canonical description). The
key idea in loop quantum gravity is the consideration of variables
with support on one-dimensional objects (loops or graphs) and this
can be achieved from either starting point.

\paragraph{An augmented algebra for quantisation.}
If we select the connection $A_a^i$ as one of the basic variables of
the theory, then it is convenient to also employ a redefined
$\xi$-momentum $\pi_i$ written as
\begin{eqnarray}
\pi_i := {}^{\xi}p_i + \tilde{E}^a_j \frac{\delta G_a^j}{\delta
\xi^i},
\end{eqnarray}
where $G_a^i = - \beta^{-1} w^i_l \partial_a \xi^l + v^i_l
\partial_a \xi^l$. The symplectic form $\Omega_1$ then reads
\begin{eqnarray}
\Omega_1 = \int dt \int d^3x \; (\delta \tilde{E}^a_i \wedge \delta
A_a^i + \delta \pi_i \wedge \delta \xi^i) \label{Omega1n}.
\end{eqnarray}

It is important to note that the variables $\xi^i, \pi_i$ commute
with the ${\cal G}_{Gauss}$ constraint. They are therefore invariant
under the gauge SU(2) rotations that characterise the $A_a^i$
connection.

 The symplectic form (\ref{Omega1n}) gives rise to the history algebra
\begin{eqnarray}
\{ A_a^i(t,x), \tilde{E}^b_j(t',x') \} &=& \delta_a^b \delta^i_j
\delta(t,t') \delta^3(x,x') \\
\{\xi^i(t,x), \pi_j(t',x') \} &=& \delta^i_j \delta(t,t')
\delta^3(x,x').
\end{eqnarray}
The corresponding canonical algebra is clearly
\begin{eqnarray}
\{ A_a^i(x), \tilde{E}^b_j(x') \} &=& \delta_a^b \delta^i_j
\delta^3(x,x') \\
\{\xi^i(x), \pi_j(x') \} &=& \delta^i_j  \delta^3(x,x').
\end{eqnarray}

This is essentially an augmentation of the canonical algebra of the
standard theory by additional canonical variables that correspond to
the internal field $n^I$ and its conjugate momentum. The subalgebra
generated by the connection and the densitised triad generates the
usual loop algebra employed in quantisation, that is defined in
terms of the $T^0$ and the $T^1$ variables.

It follows that a quantisation procedure that emphasises spacetime
covariance and  gauge-fixing independence should augment the loop
algebra by other variables, defined through $\xi^i$ and $\pi_i$.
Clearly the explicit form of the quantisation algebra will affect
the resulting quantum theory. The key issue at this point is whether
a suitable algebra and representation thereof can be found, such
that the quantum mechanical imposition of the ${\cal F'}$ constraint
will reproduce the standard constructions of loop quantum theory.
This is plausible, because the ${\cal F'}$ constraint is linear with
respect to $\pi$. However, the agreement is not {\em a priori}
guaranteed, because the result may be sensitive not only to the
representation of the additional variables, but also on the
procedure one employs in the quantum mechanical implementation of
the constraints.

One possible procedure for quantising the loop algebra--augmented by
the variables $\xi^i$ and $\pi_i$--may be provided by a
generalisation of a technique developed in Ref. \cite{IS03}. In this
reference  a histories description of quantum fields is developed,
with the particular aim to treat the foliation as a potential
quantum variable.

Finally, we would like to comment on the relation between the
results of this paper with those of Refs.\cite{samuel00a,samuel00b}.
In these papers, Samuel explains that the Barbero connection does
not have a spacetime interpretation, because it is not the pullback
of a spacetime connection onto the three-surface and because of the
gauge-fixing employed in its definition. While we fully agree with
these statements, we believe that they should not be taken to imply
that the Barbero connection is not a suitable variable for
quantisation. The Barbero connection appears naturally in the
Legendre transform of the Holst action even in absence of gauge
fixing. In the histories framework in particular, it may be employed
to  define loop variables similar to those employed in loop quantum
gravity, without compromising the spacetime symmetries of the
theory. The price we have to pay is that we need to consider
additional variables, spanning the space $\Pi_1$ (or its single-time
analogue in the canonical case) and we have to implement the
constraints ${\cal F'}$ at the quantum level. With these
modifications the loop quantisation of gravity will fully preserve
its spacetime character: in the histories framework one expects that
the resulting Hilbert space will carry a representation of the
spacetime diffeomorphism group. The outstanding issue  is whether
the ensuing formalism will lead to the same results as the standard
one.

\section{Conclusions}

We developed the histories description for classical gravity,
described in term of the Holst Lagrangian. The basic variables at
the covariant level is an $SL(2,{\bf C})$ connection and a field of
tetrads on spacetime $M$ together with their conjugate variables.
The history space $\Pi_{cov}$ that they span carries a symplectic
representation of the group $Diff(M)$ of spacetime diffeomorphisms.

The next step involves the introduction of an equivariant foliation
functional, implementing the translation from the spacetime theory
to an one-parameter family of canonical structures. The Legendre
transform of the Holst Lagrangian leads to an identification of a
set of constraints, both of first- and second-class types. The
construction of the metric-based  theory is fully transferred into
this construction and the set of constraints is invariant under the
action of the diffeomorphism group. Hence, the generators of the
spacetime diffeomorphism group can also be projected onto the
reduced state space.

The key issue then arises, which variables should be chosen for the
quantisation of the history theory. Following the spirit of loop
quantum gravity, we should try to construct a loop algebra and
identify a Hilbert space by studying the algebra's representation
theory.

The obvious place to start would be to consider the loop algebra
that corresponds to the spacetime $SL(2,{\bf C})$ connection of the
covariant description. This however would involve developing a
representation theory for loop variables with a non-compact gauge
group. Moreover, we would have to identify a new role for the tetrad
fields, because at this level they commute with the connection
variables.

It may therefore be more profitable to work with `internal' fields,
which appear as one-parameter families of the standard canonical
variables. This will have the benefit of allowing the consideration
of connections with compact gauge group. Here, however a
complication arises because of the gauge-dependence of the Barbero
connection. We showed that is possible to implement the second-class
constraints together with the primary first-class ones, thus
arriving at an intermediate state space $\Pi_1$ spanned by the
variables $\tilde{E}^a_i, \omega_a^i, \xi^i, {}^{\xi}\pi^i$, which
were defined in Section 3.2. The remaining constraints are the
Hamiltonian constraint, the three momentum constraints, the three
Gauss-law constraints  and the three boost gauge constraints. The
analogous canonical analysis proceeds by imposing a gauge-fixing
condition, which in our notation corresponds to taking $\xi^i = 0$.
This allows one to get rid of the boost gauge constraints. In the
resulting space, the variable $\omega_a^i$ is the $SU(2)$ Barbero
connection, which remains conjugate to the triad. The remaining
constraints take a simple form.

In the histories framework, however, we emphasise the general
covariance of the theory: the imposition of a gauge fixing condition
breaks the background independence of the theory. Hence, the
spacetime diffeomorphism group does not descend to the resulting
phase space.
It follows therefore that a spacetime covariant loop quantisation
that employs path variables should start at the very least from the
space $\Pi_1$. Nonetheless, a pullback of the Barbero connection is
well defined on $\Pi_1$: there exists a simple combination $A_a^i$
of the variables $\xi^i$ and $\omega_a^i$, which is (i) conjugate to
the triad and (ii) behaves as an $SU(2)$ connection, with respect to
the rotation gauge constraints. Hence a history quantisation of
$\Pi_1$ may be envisioned, that will employ variables defined with
support on a two-dimensional cylinder---giving a history analogue of
the $T_0$ variables---and a three-dimensional space $S\times R$
(where $S$ is a spatial two-surface) as a history analogue of the
$T_1$ variables.

However, these variables have to be implemented with additional ones
that involve $\xi^i$ and its conjugate momentum. This suggests that
the kinematical Hilbert space, if constructed in terms of variables
with support on graphs as in the canonical loop quantum gravity
programme, should involve at each point of the graph not only a
representation of the $SU(2)$ group but also a mathematical object
describing a unit time-like vector. This additional degree of
freedom should disappear when the  constraint ${\cal F'}$ is
implemented, but there is no {\em a priori} guarantee that the
resulting theory will be identical with the one obtained from the
usual loop quantisation: it is plausible for example that the
spectrum of physically relevant operators (e.g. the area operator)
may develop additional degeneracies.

\bigskip

\noindent{\large\bf Acknowledgements}

\noindent I would like to thank Charis Anastopoulos for very helpful
discussions, and Maria Electra Plakitsi for her inspiring
contribution throughout this work. I gratefully acknowledge support
from the EP/C517687 EPSRC grant. This work was partially supported
by a Pythagoras I grant from the Hellenic Ministry of Education.

\begin{appendix}

\section{The Legendre transform}

The $3+1$ decomposition of the Holst Lagrangian (\ref{Holst}) is
straightforward, and its expression in terms of the variables
$E_a^I$, ${}^3\omega_a^{IJ}$, $\alpha^{IJ}$, $n^I$, $N$, $N_a$ can
be found in the related literature \cite{Holst}. Our point of
departure is the substitution of the variables $\tilde{E}^a_i$,
$\omega_a^i$, $\chi _a^i$, $\mu^i$, $\nu^i$, $\tilde{N}$, $N_a$,
$\xi^i$, defined in Section 3.2. The substitution of these variables
yields---after tedious calculations---the following form for the
action
\begin{eqnarray}
S =\!\!\!\int \!\!\!dt \!\!\int \!\!d^3 x (\tilde{E}^a_i
\dot{\omega}_a^i \!-\! \e^{ijk}\tilde{E}^a_i(\chi_a^j v^k_l \!+\!
\omega_a^j w^k_l)
\partial_a \xi^l \!+\! \rho^i {\cal G}^i \!+\! \sigma_i {\cal F}^i
\!+\! N^a {\cal H}_a \!+\! \tilde{N} {\cal H})
\label{action}\hspace{0.2cm}
\end{eqnarray}
where ${\cal G}^i, {\cal F}^i, {\cal H}_a, {\cal H}$ are given by
equations (\ref{fsc1}-\ref{fsc4}). The functions $v^k_l, w^k_l$ are
defined in terms of the `boost' $\Lambda^I_J$ of Eq. (\ref{boost})
as follows
\begin{eqnarray}
\Lambda^0_K \delta \Lambda^{Ki} &=& v^i_l \delta \xi^l \\
\nonumber \Lambda^i_K \delta \Lambda^{Kj} &=& \e^{ijk} w^k_l \delta
\xi^l, \label{vw1}
\end{eqnarray}
where $\xi^i$ parameterises $n^I$ as $n^I = (\sqrt{1 + \xi^2},
\xi^i)$. Then,
\begin{eqnarray}
v^i_l &:=& -\frac{1}{\sqrt{1 +\xi^2}-1} (\delta^i_l - \frac{\xi^i
\xi_l}{\sqrt{1 + \xi^2}}) \nonumber \\
w^i_l &:=& \frac{1}{\sqrt{1 +\xi^2}-1} \; \e^{i}{}{}_{jl} \xi^j.
\label{vw2}
\end{eqnarray}

Finally we perform the Legendre transform of the action
(\ref{action}). The primary constraints (\ref{1sc}-\ref{pf4}) are
immediately determined. The secondary constraints (\ref{fsc1},
\ref{fsc4}) are also  straightforward to read. The non-trivial part
is the identification of the constraint (\ref{5sc}). This is
obtained from the Poisson bracket of the constraint (\ref{2sc}) with
the Hamiltonian, since the variables $\chi_a^i$ play the role of
`Lagrange multipliers' in the action (\ref{action}). For this
purpose it is convenient to define the $3 \times 3$ matrix
$\tilde{E}^{ai} \chi_a^j$ and to split it into a symmetric part
$M^{ij}$ and an antisymmetric part $\e^{ijk} L_k$ defined as
$\tilde{E}^{ai} \chi_a^j := M^{ij} + \e^{ijk}L_k$.

The antisymmetric part corresponding to $L_k$ is determined from the
constraint (\ref{4sc}). We then only need to vary the Hamiltonian
with respect to the symmetric part $M^{ij}$. We observe that the
only part of the Hamiltonian in which $M^{ij}$ appears explicitly is
the Hamiltonian constraint ${\cal H}$. Variation of the Hamiltonian
constraint yields
\begin{eqnarray}
M^{kl} = U^{kl} + (1+\beta^{-2}) \tilde{\Gamma}^{kl},
\end{eqnarray}
where $U^{kl}$ is the symmetric part of the matrix
\begin{eqnarray}
\tilde{E}^{ak}(\beta^{-1} \omega_a^l + (1+\beta^{-2}) w^n_l
\partial_a\xi^l),
\end{eqnarray}
and where
\begin{eqnarray}
\tilde{\Gamma}^{kl} := \tilde{E}^a_i \tilde{E}^b_j \e^{ijk}
\partial_{[a}\tilde{E}_{b]}^l - \frac{1}{2} \delta^{kl} \tilde{E}
^a_i \tilde{E}^b_j \e^{ijm}\partial_{[a}\tilde{E}_{b]m}.
\end{eqnarray}
This constraint enables us to determine explicitly $\chi_a^i$ in
terms of the remaining canonical variables
\begin{eqnarray}
\chi_a^i = \beta^{-1} \omega_a^i + (1+ \beta^{-2}) \left[\Gamma_a^i
+ \tilde{E}_{aj} \tilde{E}^{bi} w^j_l
\partial_b \xi^l +  \e^{ijk} \tilde{E}_a^j {\cal
G}^k \right].
\end{eqnarray}
The variable $\Gamma_a^i = \tilde{E}_{ak} \tilde{\Gamma}^{ki} -
\e^{ijk} \tilde{E}_{aj} \partial_b \tilde{E}^{bk}$ is the unique
torsionless connection corresponding to the  triad $E_a^i$.

If we solve the  second-class constraints and then choose the gauge
condition $\xi^i = 0$---thus solving three of the gauge
constraints---we find that on the constraint surface
\begin{eqnarray}
 \chi_a^i = \beta^{-1} \omega_a^i + (1+\beta^{-2}) \Gamma_a^i.
\end{eqnarray}
The expression above allows the elimination of the $\chi_a^i$
variables from our description. As shown in Refs. \cite{Holst,
BeS01} it leads to the standard expression for the constraints in
the temporal gauge.

\end{appendix}
\end{document}